\documentclass[showpacs,aps,prb,twocolumn,superscriptaddress]{revtex4}
\usepackage{amsmath,amssymb}
\usepackage[final]{graphicx}
\usepackage{stmaryrd}
\usepackage{wasysym}
\usepackage{bm}

\begin{document}

\title{Spectral features due to inter-Landau-level transitions in the Raman spectrum of bilayer graphene}
\author{Marcin Mucha-Kruczy\'{n}ski}
\affiliation{Department of Physics, Lancaster University, Lancaster, LA1~4YB, United Kingdom}
\author{Oleksiy Kashuba}
\affiliation{Department of Physics, Lancaster University, Lancaster, LA1~4YB, United Kingdom}
\affiliation{Institute for Theoretical Physics A, RWTH Aachen, D-52074 Aachen, Germany}
\author{Vladimir I. Fal'ko}
\affiliation{Department of Physics, Lancaster University, Lancaster, LA1~4YB, United Kingdom}

\begin{abstract}

We investigate the contribution of the low-energy electronic excitations towards the Raman spectrum of bilayer graphene for the incoming photon energy $\Omega\gtrsim 1$eV. Starting with the four-band tight-binding model, we derive an effective scattering amplitude that can be incorporated into the commonly used two-band approximation. Due to the influence of the high-energy bands, this effective scattering amplitude is different from the contact interaction amplitude obtained within the two-band model alone. We then calculate the spectral density of the inelastic light scattering accompanied by the excitation of electron-hole pairs in bilayer graphene. In the absence of a magnetic field, due to the parabolic dispersion of the low-energy bands in a bilayer crystal, this contribution is constant and in doped structures has a threshold at twice the Fermi energy. In an external magnetic field, the dominant Raman-active modes are the $n^{-}\!\rightarrow\! n^{+}$ inter-Landau-level transitions with crossed polarisation of in/out photons. We estimate the quantum efficiency of a single $n^{-}\!\rightarrow\! n^{+}$ transition  in the magnetic field of 10T as $I_{n^{-}\rightarrow n^{+}}\sim 10^{-12}$.

\end{abstract}

\pacs{73.22.Pr, 71.70.Di, 78.67.Wj, 81.05.ue}
\maketitle

\section{introduction}

Bilayer graphene \cite{mccann_prl_2006, geim_novoselov_review_2007} is a representative of the recently discovered family of new carbon allotropes.\cite{geim_novoselov_review_2007} It attracted attention by the observation of an unusual sequencing of plateaus in the quantum Hall effect \cite{mccann_prl_2006, geim_novoselov_review_2007} and the possibility to modify its spectrum by opening a small gap and induce an insulating state with an external electric field.\cite{geim_novoselov_review_2007, mccann_prb_2006, oostinga_nature_materials_2007} The electronic properties of bilayer graphene have been characterised using angle-resolved photoemission spectroscopy \cite{ohta_science_2006, mucha-kruczynski_prb_2008} and optical absorption in the visible \cite{nair_science_2008, gaskell_apl_2009} and infrared \cite{zhang_prb_2008, kuzmenko_prb_2009a, kuzmenko_prb_2009b, li_prl_2009, mak_prl_2009, kuzmenko_prl_2009} spectral range. Bilayer graphene has also been studied using inelastic scattering of light, subject to the detection of Raman-active lattice vibrations in this two-dimensional crystal.\cite{ferrari_prl_2006, gupta_nano_letters_2006, graf_nano_letters_2007, yan_prl_2008, malard_prl_2008, das_prb_2009, malard_prb_2007, ni_prb_2009, mafra_prb_2009} However, no theoretical or experimental study has yet been performed, on the contribution of electronic excitations towards Raman spectra of bilayer graphene. This paper presents a theory of such a contribution.

The electronic Raman spectroscopy can provide information about various single particle and collective electron excitations in the system studied. In semiconductors, it has been, for example, employed to investigate donor and acceptor states, plasmons and spin-density fluctuations involving electron spin-flip due to the spin-orbit interaction.\cite{abstreiter_springer_1984, klein_springer_1983} The inelastic scattering of photons on electrons in semiconductor placed in an external magnetic field was first discussed in Ref. \onlinecite{wolff_prl_1966}, where it was pointed out that nonparabolicity of the electronic bands is crucial for the electron-photon interaction matrix elements not to vanish. The features corresponding to the electronic contribution to the Raman scattering in an external magnetic field were observed in many semiconductors, for example, InSb (Ref. \onlinecite{slusher_prl_1967}) and GaAs.\cite{patel_prl_1968}

Recently, the Raman spectroscopy of electronic excitations in monolayer graphene has been investigated theoretically.\cite{kashuba_prb_2009} It has been shown that at high magnetic fields the inelastic light scattering accompanied by the excitation of the electronic mode with the highest quantum efficiency involves the generation of inter-band electron-hole pairs. At high (quantizing) magnetic fields this leads to the electron excitations from the Landau level (LL) $n^{-}$ at the energy $-\sqrt{2n}\hbar v/\lambda_{B}$ in the valence band to the Landau level $n^{+}$ at the energy $\sqrt{2n}\hbar v/\lambda_{B}$ in the conduction band with energies $\omega_{n}=2\sqrt{2n}\hbar v/\lambda_{B}$ and crossed polarisation of in/out photons, in contrast to the $\Delta n=\pm 1$ transitions between Landau levels which are dominant in the absorption of left and right-handed circularly polarised infrared photons.\cite{abergel_prb_2007} Raman spectroscopy, therefore, provides data supplementary to that obtained in optical absorption. This fact could be of interest in particular for bilayer graphene placed in an external magnetic field. Recent measurement of the  infrared absorption in fields up to $20$T showed \cite{henriksen_prl_2008} deviations from the tight-binding model for an ideal bilayer. Here, we offer a theory for an alternative experimental probe, with different selection rules for the inter-Landau-level excitations, which could bring some new insight into the properties of the bilayer graphene.

We study the Raman spectroscopy of electronic excitations in bilayer graphene both with and without an external magnetic field using the tight-binding approach. First, we describe processes in which after the inelastic scattering of the optical photon, an electron-hole pair is created in the low-energy bands of the bilayer. Then, we evaluate the scattering amplitude corresponding to such a process within the framework of the four-band tight-binding model and extract an effective scaterring amplitude that is incorporated into the two-band low-energy Hamiltonian for bilayer graphene.\cite{mccann_prl_2006} This effective scattering amplitude contains the influence of the high-energy bands and is different from the scattering amplitude found using only the approximate low-energy Hamiltonian. Next, we find the low-energy electronic contribution to the Raman spectra and obtain an analytic description for the selection rules and intensity of such Raman scattering in a bilayer placed in a strong (quantizing) magnetic field.

\section{Theory of the inelastic light scattering in bilayer graphene}

Bilayer graphene consists of two coupled sheets of graphene with $AB$ (Bernal) stacking characteristic of bulk graphite,\cite{dresselhaus_adv_physics_1981} Fig. 1(a). The unit cell contains four inequivalent atoms $A1$, $B1$, $A2$ and $B2$ where letters $A$ and $B$ denote two sublattices in the same layer and $1/2$ stands for the bottom/top layer. The Fermi level in graphene lies in the vicinities of the corners of the hexagonal Brillouin zone (also called valleys) known as $K_{+}$ and $K_{-}$ (Fig. 1(b)). The conventional tight-binding Hamiltonian based on $\pi$-orbitals of carbon atoms (one per atom, four in the unit cell) and expanded in momentum around the valleys reads 
\begin{equation}\label{4b_Hamiltonian}
{\hat{H}_{0}} \!= \!\left(
\begin{array}{cc}
\xi v_{3}(\sigma_{x}p_{x}-\sigma_{y}p_{y}) & \xi v \bm{\sigma}\!\cdot\!\mathbf{p} \\
\xi v \bm{\sigma}\!\cdot\!\mathbf{p} & \gamma_{1}\sigma_{x}
\end{array}\right).
\end{equation}
Here, $\bm{\sigma}=(\sigma_{x},\sigma_{y})$ and $\sigma_{x}$,$\sigma_{y}$,$\sigma_{z}$ are the Pauli matrices, electron momentum $\mathbf{p}$ is measured from the center of the valley, $v\sim 10^{6}$m/s (Refs. \onlinecite{kuzmenko_prb_2009b, li_prl_2009, malard_prb_2007, mafra_prb_2009}) is a parameter related to the nearest neighbour intralayer coupling $\gamma_{0}\sim 3$eV,\cite{kuzmenko_prb_2009b, malard_prb_2007, mafra_prb_2009}, $\gamma_{1}\sim 0.4$eV (Refs. \onlinecite{ohta_science_2006, zhang_prb_2008, kuzmenko_prb_2009a, kuzmenko_prb_2009b, li_prl_2009, malard_prb_2007, yan_prl_2008, das_prb_2009, mafra_prb_2009}) is the direct interlayer coupling, $v_{3}$ is related to the weak direct $A1\leftrightarrow B2$ interlayer hops [$v_{3}/v\sim 0.1$ (Refs. \onlinecite{mccann_prl_2006, malard_prb_2007, mafra_prb_2009})] and $\xi=\pm$ is the valley index. The basis is constructed using components corresponding to atomic sites $A1,B2,A2,B1$ in the valley $K_{+}$ and $B2,A1,B1,A2$ in $K_{-}$. One can also take into account terms quadratic in the electron momentum $\mathbf{p}$:
\begin{equation}
\delta\!{\hat{H}} \!=\! \mu \!\!\left(
\!\!\begin{array}{cc}
\frac{v_{3}}{v}\![\sigma_{\!x}(p_{x}^{2}\!-\!p_{y}^{2}) \!+\! 2\sigma_{\!y}p_{x}p_{y}] & \sigma_{x}(p_{x}^{2}\!-\!p_{y}^{2}) \!-\! 2\sigma_{y}p_{x}p_{y} \\
\sigma_{x}(p_{x}^{2}\!-\!p_{y}^{2}) \!-\! 2\sigma_{y}p_{x}p_{y} & 0 
\end{array}\!\!
\right)\!, \nonumber
\end{equation}
where $\mu=-\frac{v^{2}}{6\gamma_{0}}$. However, the influence of the $\delta\!{\hat{H}}$ term on the results of the Raman spectra analysis is negligibly small, as shown in the Appendix. 

\begin{figure}[tbp]
\centering
\includegraphics[width=.9\columnwidth]{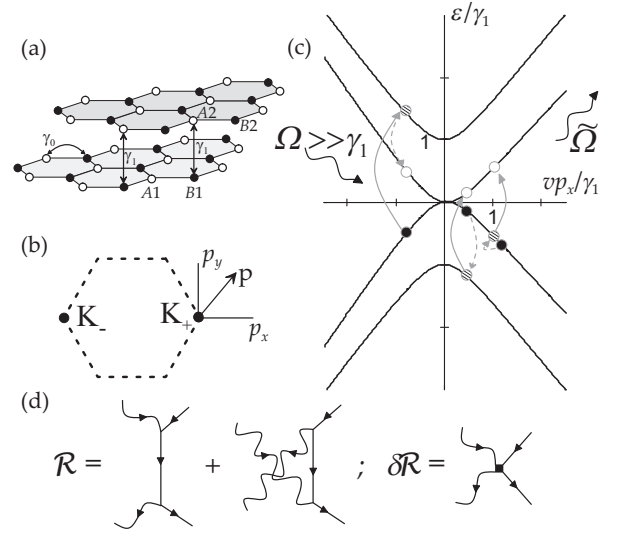}
\caption{(a) Schematic of the bilayer graphene crystal lattice.  (b) The Brillouin zone of bilayer graphene with two inequivalent valleys denoted as $K_{+}$ and $K_{-}$. (c) The band structure of bilayer graphene in the vicinity of the $K_{+}$ point along the $p_{x}$ axis. Also shown are some of the two-step processes leading to the creation of an electron-hole pair in the low-energy bands accompanied by the absorption of a photon followed by emission. Gray solid (dashed) lines indicate the first (second) step of the process. The black (white) circle denotes the hole (electron) in the final electron-hole pair, while the hatched circle represents the intermediate virtual state. Note that for any intermediate state $|\nu\rangle$ with energy $\epsilon_{\nu}$, $\Omega,\tilde{\Omega}\gg\epsilon_{\nu}$. (d) Diagrammatic representation of the scattering amplitudes $\mathcal{R}$, discussed in Sec. II, and $\delta\!\mathcal{R}$ (discussed in Appendix).}
\label{f1}
\end{figure}

The part of the resulting electronic dispersion relevant for the Raman scattering of photons with energies $\Omega<2.5$eV, is illustrated in Fig. 1(c) for the valley $K_{+}$. Two bands, later referred to as low-energy ones, touch each other at the neutrality point - the position of the Fermi energy in the neutral structure. Two other, referred to as high-energy bands, are split by the interlayer coupling, $\gamma_{1}$, from the neutrality point. The $v_{3}$ parameter leads to the trigonal warping of the electronic dispersion. Its influence is most important for very low energies, $\epsilon<5$meV.

To describe the process of inelastic scattering of light on electrons, we consider an experimental setup in which incoming laser light of energy $\Omega\gg\gamma_{1}$, in-plane momentum $\mathbf{q}$ (out-of-plane component of momentum equal to $q_{z}=\sqrt{\Omega^{2}/c^{2}-\mathbf{q}^{2}}$) and polarisation $\mathbf{l}$ is shined onto to the sample. Scattered photon has polarisation $\mathbf{\tilde{l}}$, in-plane momentum $\tilde{\mathbf{q}}$ and energy $\tilde{\Omega} = \Omega - \omega$, where $\omega$ is the Raman shift.  We also assume the temperature $T$ to be smaller than the Raman shift, $k_{B}T<\omega$ ($k_{B}$ is the Boltzmann's constant). In our case, the inelastic light scattering may occur via a one-step process (so called contact interaction) or a two-step process involving an intermediate state. The two-step process, such as shown in Fig. 1(c), involves: the absorption (or emission) of a photon with energy $\Omega$ ($\tilde{\Omega}$) transferring an electron with momentum $\mathbf{p}$ from an occupied state in the valence band into a virtual intermediate state (energy is not conserved at this stage), followed by another electron emission (or absorption) of the second photon with energy $\tilde{\Omega}$ ($\Omega$). The one-step process is the usual inelastic scattering of an incoming photon on an electron with transfer of energy to the latter. As a result of both one and two-step processes, an electron-hole pair in the low-energy bands is created with the electron and the hole having almost the same momentum ($\mathbf{p}+\mathbf{q}-\tilde{\mathbf{q}}$ and $\mathbf{p}$, respectively), since $\mathbf{q},\tilde{\mathbf{q}}\ll\mathbf{p}$ and the momentum transfer from light is negligible ($v/c\sim 3\cdot 10^{-3}$). Therefore, $\mathbf{p}+\mathbf{q}-\tilde{\mathbf{q}}\approx\mathbf{p}$ and due to the approximately electron-hole symmetric band structure in the vicinity of Brillouin zone corners, the electron initial and final energies $\epsilon_{i}$ and $\epsilon_{f}$ are related, $\epsilon_{f}\approx-\epsilon_{i}$. 

To include the interaction of the electrons with photons, we construct the canonical momentum $[\mathbf{p}-e(\mathbf{A}(\mathbf{r},t')+\tilde{\mathbf{A}}(\mathbf{r},t''))]$, where $\mathbf{A}(\mathbf{r},t')$ and $\tilde{\mathbf{A}}(\mathbf{r},t'')$ are the vector potentials of the incoming and outgoing light, respectively,
\begin{equation}\begin{split}
& \mathbf{A}(\mathbf{r},t')=\frac{1}{\sqrt{2\epsilon_{0}\Omega}} \left( \mathbf{l}e^{i(\mathbf{q}\cdot\mathbf{r}-\Omega t')/\hbar} b_{\mathbf{q},q_{z},\mathbf{l}} + h.c \right); \\
& \tilde{\mathbf{A}}(\mathbf{r},t'')=\frac{1}{\sqrt{2\epsilon_{0}\tilde{\Omega}}} \left( \tilde{\mathbf{l}}^{*}e^{-i(\tilde{\mathbf{q}}\cdot\mathbf{r}-\tilde{\Omega}t'')/\hbar} b^{\dagger}_{\tilde{\mathbf{q}},\tilde{q}_{z},\tilde{\mathbf{l}}} + h.c. \right);
\end{split}\end{equation}
and $b_{\mathbf{q},q_{z},\mathbf{l}}$ is an annihilation operator for a photon with in-plane momentum $\mathbf{q}$, out-of-plane momentum component $q_{z}$ and polarisation $\mathbf{l}$. We expand the resulting Hamiltonian up to the second order in the vector potential and write down the interaction part,
\begin{equation}\label{int_Hamiltonian}
{\hat{H}_{\textrm{int}}} = \mathbf{j}\!\cdot\!\big( \mathbf{A}(\mathbf{r},t')+\tilde{\mathbf{A}}(\mathbf{r},t'') \big) + \frac{e^{2}}{2}\sum_{i,j}\frac{\partial^{2}\hat{H}_{0}}{\partial p_{i} \partial p_{j}}A_{i}\tilde{A}_{j},
\end{equation}
where $\mathbf{j}=-e\frac{\partial\hat{H}_{0}}{\partial\mathbf{p}}$ is the current vertex.

We aim to calculate the spectral density $g(\omega)$ and the quantum efficiency (intensity) of the Raman scattering, $I$. The quantum efficiency describes the ratio of the flux of outgoing, inelastically scattered photons to the flux of the incoming photons, and is an integral, $I=\int \!\!d\omega g(\omega)$, of the spectral density $g(\omega)$ representing the probability for the incoming photon to scatter inelastically with energy $\tilde{\Omega}=\Omega-\omega$, where $\omega$ is the Raman shift.

The quantum efficiency expresses the total probability for single incoming photon to scatter inelastically in a proccess under consideration - that is, to scatter on an electron and excite an electron-hole pair in the low-energy bands. The probability for the incoming photon to scatter with the Raman shift $\omega$ in a particular direction (defined by the momentum $\tilde{\mathbf{q}}$ of the scattered photon), is, in turn, characterised by the angle-resolved probability of scattering $w(\tilde{\mathbf{q}})$. Finally, the scattering probability $w(\tilde{\mathbf{q}})$ that one photon is scattered with the excitation of an electron-hole (e-h) pair in the final state is related, as $w\propto|\mathcal{R}|^{2}$, to the scattering amplitude $\mathcal{R}$ of the Raman process. 

The amplitude $\mathcal{R}$ is the sum of the amplitudes corresponding to the one-step and two-step processes. Only terms quadratic in electron momentum $\mathbf{p}$ which appear in the addition $\delta\!{\hat{H}}$ to the Hamiltonian in (\ref{4b_Hamiltonian}), contribute to the contact interaction. We show in Appendix that this contribution is much smaller than the leading contribution from the two-step processes, thus, we neglect it in further considerations. To find $\mathcal{R}$, illustrated using Feynman diagrams shown in Fig. 1(d), we describe a two-step transition which involves an intermediate virtual state $|\nu\rangle$ with energy $\epsilon_{\nu}$, as\begin{widetext}
\begin{align}\begin{split}\label{2nd_order_perturbation}
\mathcal{R} & = - \frac{1}{2\epsilon_{0}\sqrt{\Omega\tilde{\Omega}}} \sum_{\nu}\int_{-\infty}^{\infty}\int_{-\infty}^{t'} e^{\frac{i}{\hbar}\left( \epsilon_{f}-\epsilon_{\nu} \right)t'} \big(\mathbf{j}\!\cdot\!\tilde{\mathbf{l}}^{*}\big) e^{-\frac{i}{\hbar}(\tilde{\mathbf{q}}\cdot\mathbf{r}-\tilde{\Omega}t')} |\nu\rangle\langle\nu| e^{\frac{i}{\hbar}(\mathbf{q}\cdot\mathbf{r}-\Omega t'')} \big(\mathbf{j}\!\cdot\!\mathbf{l}\big) e^{\frac{i}{\hbar}\left( \epsilon_{\nu} - \epsilon_{i} \right)t''} dt' dt'' - \\
& - \frac{1}{2\epsilon_{0}\sqrt{\Omega\tilde{\Omega}}} \sum_{\nu}\int_{-\infty}^{\infty}\int_{-\infty}^{t'} e^{\frac{i}{\hbar}\left( \epsilon_{f}-\epsilon_{\nu} \right)t'} \big(\mathbf{j}\!\cdot\!\mathbf{l}\big) e^{\frac{i}{\hbar}(\mathbf{q}\cdot\mathbf{r}-\Omega t')} |\nu\rangle\langle\nu| e^{-\frac{i}{\hbar}(\tilde{\mathbf{q}}\cdot\mathbf{r}-\tilde{\Omega}t'')} \big(\mathbf{j}\!\cdot\!\tilde{\mathbf{l}}^{*}\big) e^{\frac{i}{\hbar}\left( \epsilon_{\nu} -\epsilon_{i} \right)t''} dt' dt''.
\end{split}\end{align}\end{widetext}
The virtual state $|\nu\rangle$ may belong to any of the four bands, since an electron is excited from a state with momentum $\mathbf{p}$ to a state with momentum $\mathbf{p}+\mathbf{q}$ or $\mathbf{p}-\tilde{\mathbf{q}}$ depending on the accompanying photon process. At this step of the calculation we still work with the four-band Hamiltonian (\ref{4b_Hamiltonian}), to include the influence of the high-energy "split" bands. In Eq. (\ref{2nd_order_perturbation}), the first (second) term corresponds to processes in which the photon is absorbed (emitted) in the first step and emitted (absorbed) in the second step of the process and is given by the first (second) diagram in the expression for $\mathcal{R}$ in Fig. 1(d). Integration in the time-dependent perturbation theory in Eq. (\ref{2nd_order_perturbation}) can be performed by changing variables to $\tau=t'-t''$, which varies at the scale of $\omega^{-1}$, $\omega=\Omega-\tilde{\Omega}$, and $\bar{t}=(t'+t'')/2$, which varies at the scale of ${\bar{\Omega}}^{-1}$, $\bar{\Omega}=(\Omega+\Omega ')/2\gg\omega$. For incoming and outgoing photons, $\Omega,\tilde{\Omega}\gg\gamma_{1}$, and we also study the low energy excitations in the final states with $\omega\ll\gamma_{1}$. This allows us to expand factors $\frac{1}{\pm \bar{\Omega}-\epsilon_{\nu}}$ resulting from the integration over $\tau$ in powers of $\left( \epsilon_{\nu}/\Omega \right)$, keeping terms of the order of $1$ and $\left(\gamma_{1}/\Omega\right)$ [the latter appear when the virtual state is taken to be in the high-energy bands] and to perform summation over the intermediate virtual states of the process. Consequently, the amplitude $\mathcal{R}$ takes the form of a matrix
\begin{align}
& \mathcal{R} \!\approx\! \frac{e^{2}\hbar^{2}v^{2}}{\epsilon_{0}\Omega^{2}} \!\left\{ \!-i \!\left( \begin{array}{cc}
\sigma_{z} & 0 \\
0 & \sigma_{z} 
\end{array} \right) \!\!\left( \mathbf{l}\times\tilde{\mathbf{l}}^{*} \right)_{z} \!+ \!\frac{\bm{\mathcal{M}}\!\cdot\!\mathbf{d}}{\Omega}  \right\}\!\delta\!\!\left( \epsilon_{f} \!-\! \epsilon_{i} \!-\! \omega \right)\!; \nonumber \\
& \mathbf{d}=(l_{x}\tilde{l}_{y}^{*}+l_{y}\tilde{l}_{x}^{*},l_{x}\tilde{l}_{x}^{*}-l_{y}\tilde{l}_{y}^{*});~~ \bm{\mathcal{M}}=(\mathcal{M}_{x},\mathcal{M}_{y}); \nonumber\\ & 
\mathcal{M}_{x} \!=\! \left( \begin{array}{cc}
\gamma_{1}\sigma_{y} & \xi v \!\left( \sigma_{y}p_{x}+\sigma_{x}p_{y} \right) \\
\xi v \!\left( \sigma_{y}p_{x}+\sigma_{x}p_{y} \right) & 0
\end{array} \right); \nonumber\\ & 
\mathcal{M}_{y} \!=\! \left( \begin{array}{cc}
\gamma_{1}\sigma_{x} & \xi v \!\left( \sigma_{x}p_{x}-\sigma_{y}p_{y} \right) \\
\xi v \!\left( \sigma_{x}p_{x}-\sigma_{y}p_{y} \right) & 0
\end{array} \right). \nonumber
\end{align}

Below, we analyse the contribution of electronic modes toward the low-energy part of Raman spectrum with the photon energy shift $\omega<\gamma_{1}/2$, which is determined by the excitation of the electron-hole pairs in the low-energy (degenerate) bands with $vp\ll\gamma_{1}$. At such low energies, the band structure as well as Landau level structure can be described by the effective two-band Hamiltonian written in the basis of orbitals on the sites $A1$ and $B2$,\cite{mccann_prl_2006}
\begin{equation}\label{2b_Hamiltonian}
{\hat{H}}_{\textrm{eff}} \! = \!-\frac{v^{2}}{\gamma_{1}}\left[\left( p_{x}^{2} - p_{y}^{2} \right)\sigma_{x} + 2p_{x}p_{y}\sigma_{y} \right].
\end{equation}
To describe the excitation of the low-energy modes corresponding to the transitions between low-energy band states described by ${\hat{H}}_{\textrm{eff}}$, we take only the part of $\mathcal{R}$ which acts in that two-dimensional Hilbert space, keep terms in the lowest relevant order in $vp/\gamma_{1}\ll 1$ and $\gamma_{1}/\Omega\ll 1$, and write down an effective amplitude $\mathcal{R}_{\textrm{eff}}$,
\begin{equation}\label{eff_amp}
\mathcal{R}_{\textrm{eff}}\approx\!\frac{e^{2}\hbar^{2}v^{2}}{\epsilon_{0}\Omega^{2}}\! \left\{\!
 -i\sigma_{z} \!\big( \mathbf{l}\times\tilde{\mathbf{l}}^{*} \big)_{z}
 \!+\! \frac{\gamma_{1}}{\Omega} \left[ \sigma_{x}d_{y} + \sigma_{y}d_{x} \right] \!
\right\}\!.
\end{equation}

We point out that the above matrix cannot be obtained within a theory constrained by the two-band approximation, Eq. (\ref{2b_Hamiltonian}), from the very beginning. Seemingly, one may try to define a contact-interaction-like term due to the terms quadratic in the electron momentum $\mathbf{p}$ in Eq. (\ref{2b_Hamiltonian}), which carries a prefactor $\frac{e^{2}\hbar^{2}v^{2}}{\epsilon_{0}\gamma_{1}\Omega}$, which may suggest a greater magnitude of scattering than prefactor $\frac{e^{2}\hbar^{2}v^{2}}{\epsilon_{0}\Omega^{2}}$ above. However, the scattering amplitude obtained within this model can only be applied to photons with $\Omega<\gamma_{1}$, which is hardly relevant for Raman spectroscopy since the latter is usually performed with laser beams using $\Omega\sim 1.3-2.8$eV. \cite{ferrari_prl_2006, gupta_nano_letters_2006, graf_nano_letters_2007, malard_prb_2007, yan_prl_2008, malard_prl_2008, das_prb_2009, ni_prb_2009, mafra_prb_2009}

The angle-resolved probability of the Raman scattering, $w(\tilde{\mathbf{q}}\approx \mathbf{0})$, determined using Fermi's golden rule and with the help of Eq. (\ref{eff_amp}), is
\begin{equation}
w=\frac{2}{c\pi\hbar^{3}}\int\! d\mathbf{p} |\langle f|\mathcal{R}_{\textrm{eff}}|i\rangle|^{2} \times f_{i}\left( 1-f_{f} \right)\delta\!\!\left( \epsilon_{i} + \omega -\epsilon_{f} \right), \nonumber
\end{equation}
where $f_{i}$ and $f_{f}$ are filling factors of the initial and final electronic state, respectively, and the spin and valley degeneracies have already been taken into account. This gives \cite{footnote_v3}
\begin{align}
& w \approx\! 
\frac{\gamma_{1}e^{4}\hbar v^{2}}{c\epsilon_{0}^{2}\Omega^{4}}\left\{ \Xi_{s} + \frac{\gamma_{1}^{2}}{2\Omega^{2}} \Xi_{o} \right\}\theta(\omega-2\mu); \\
& \Xi_{s}=\left|\mathbf{l}\times\tilde{\mathbf{l}}^{*}\right|^{2},~ \Xi_{o}=1+\big(\mathbf{l}\times{\mathbf{l}}^{*}\big)\cdot\big(\tilde{\mathbf{l}}\times\tilde{\mathbf{l}}^{*}\big). \nonumber
\end{align}
Above, the first term with polarization factor $\Xi_{s}$ describes the contribution of photons scattered with the same circular polarization as the incoming beam. The second term, with polarization factor $\Xi_{o}$, represents the scattered photons with circular polarization opposite to the incoming beam.

In turn, the angle-integrated spectral density of Raman scattering $g(\omega)$ is
\begin{align}\label{spectral_density}
g(\omega) & = \iint \frac{d\tilde{\mathbf{q}}d\tilde{q}_{z}}{(2\pi\hbar)^3}\, w\, \delta\!\! \left( \tilde{\Omega} - c\sqrt{\tilde{\mathbf{q}}^{2}+{\tilde{q}_{z}}^{2}} \right) \\ & = 2\left(\frac{e^{2}}{4\pi\epsilon_{0}\hbar c}\frac{v}{c}\right)^{2}\frac{\gamma_{1}}{\Omega^{2}}
\left\{ 2 \Xi_{s} + \frac{\gamma_{1}^{2}}{\Omega^{2}} \Xi_{o} \right\}\theta(\omega-2\mu). \nonumber 
\end{align}
Here, the constant spectral density $g$ as a function of $\omega$ reflects the parabolicity of the low-energy bands and thus, energy-independent density of states in the bilayer. This is different in monolayer graphene, where $g(\omega)\propto\omega$, reflecting the energy-dependent density of states of electron-hole pairs.\cite{kashuba_prb_2009} The characteristic of monolayer graphene crossed polarisation of in/out photons is retained in the case of the bilayer system. Experimentally, constant spectral density $g$ in undoped bilayer graphene is impossible to distinguish from a homogeneous background. However, if the chemical potential $\mu$ is not at the neutrality point, then transitions with $\omega<2\mu$ are essentially blocked. Although new processes, resulting in the creation of the intraband electron-hole pair excitations and very small $\omega$, are possible for $\mu\neq 0$, their contribution carries additional prefactor $v/c\sim\frac{1}{300}$. \cite{wolff_prl_1966} Explicit calculation performed for the monolayer graphene showed that the quantum efficiency of the intraband transitions was of the order of $10^{-15}$. \cite{kashuba_prb_2009} In contrast, for chemical potential $\mu\sim 50$meV (corresponding to additional carrier density $n_{0}\sim 1.5\times 10^{12}$cm$^{-2}$), the lost quantum efficiency due to the blocked interband transitions is, according to Eq. (\ref{spectral_density}), $\Delta I\sim 10^{-12}$.

\section{Inter-Landau-level transitions in bilayer graphene Raman}

The quantization of electron states into Landau levels gives the Raman spectrum due to the electronic excitations a pronounced structure which can be used to detect their contribution experimentally. We only consider here low-energy Landau levels, as at high energies the Landau level broadening due to, for example, electron-phonon interaction, will smear out the LL spectrum. In strong magnetic fields, low-energy Landau levels are sufficiently described \cite{abergel_epjst_2007} by
\begin{align}
& \epsilon_{n^{\alpha}}=\alpha\frac{2\hbar^{2}v^{2}}{\gamma_{1}\lambda_{B}^{2}}\sqrt{n(n-1)}; \\ & \Psi_{n^{\alpha}} = \!\left(\!\begin{array}{c}
\psi_{n} \\ 0
\end{array}\!\right) , \, n=0,1;\,\, \Psi_{n^{\alpha}} = \!\frac{1}{\sqrt{2}}\!\left(\!\begin{array}{c}
\psi_{n} \\ \alpha\psi_{n-2}
\end{array}\!\right) , \, n\geq 2; \nonumber
\end{align}
where $\lambda_{B}=\sqrt{\hbar/eB}$ is the magnetic length, $n$ is the Landau level index and $\alpha=+$ denotes the conduction and $\alpha=-$ the valence band. Also, $\psi_{n}$ is the normalised $n$-th Landau level wavefunction. In a neutral bilayer, all LLs have additional fourfold degeneracy (two due to the electron spin and two due to the valley). Moreover, levels $n=0$ and $n=1$ are degenerate at $\epsilon=0$ giving rise to an 8-fold degenerate LL. We can project our effective transition amplitude $\mathcal{R}_{\textrm{eff}}$ onto the eigenstates $\Psi_{n^{\alpha}}$ to find the electronic Raman spectrum in the presence of a strong external magnetic field. This leads to the following selection rules for allowed electronic transitions from the initial level $n^{-}$:
\begin{equation}\label{sel_rules}
i)\, n^{-}\!\rightarrow\! n^{+};\,\,\,\,\,\,\,\,\,\,\, ii)\, (n\mp 1)^{-}\!\rightarrow\! {(n\pm 1)}^{+}.
\end{equation}
\begin{figure}[tbp]
\centering
\includegraphics[width=.95\columnwidth]{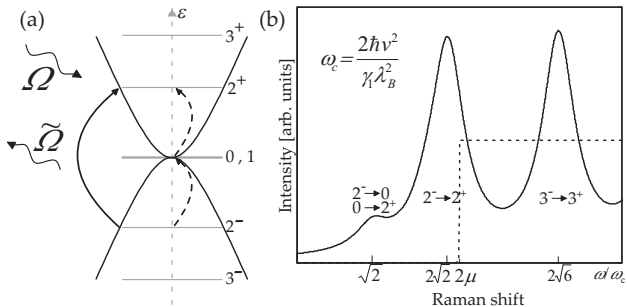}
\caption{(a) Schematic of allowed inter-LL transitions accompanying the Raman scattering. The solid (dashed) line represents the first dominant (weaker) transition $2^{-}\!\leftarrow\!2^{+}$ (pair $2^{-}\!\leftarrow\!0$ and $0\!\leftarrow\!2^{+}$). (b) The low-energy electronic contribution to the Raman spectrum in bilayer graphene. The solid (dashed) line represents the spectrum in the presence (absence) of an external magnetic field and chemical potential $\mu=0$ ($\mu\neq 0$). For the spectrum in a magnetic field, corresponding inter-LL transitions have been attributed to each peak.}
\label{f2}
\end{figure}
Among those, $i)$ is the dominant transition. These selection rules, represented schematically in Fig. 2(a), show that using Raman spectroscopy, one can probe different electronic excitations than in optical spectroscopy, where the selection rules are $\Delta n= \pm 1 $.\cite{abergel_prb_2007, mucha-kruczynski_jpcm_2009} For a neutral bilayer, the angle-integrated spectral density $g(\omega)$ of Raman scattering in the magnetic field is equal to:\begin{widetext}
\begin{align}\label{raman_mag_field}
g(\omega) & \approx 16 \,\Xi_{s} \!\left( \frac{e^{2}}{4\pi\epsilon_{0}\hbar c} \frac{v}{c} \right)^{2}\! \left(\frac{\hbar v}{\lambda_{B}\Omega}\right)^{2} \sum_{n\geq 2}\gamma(\omega - 2\epsilon_{n^{+}}) + \delta\!g(\omega); \\ 
\delta\!g(\omega) & = 8 \,\Xi_{o} \!\left( \frac{\gamma_{1}}{\Omega} \right)^{2} \!\left( \frac{e^{2}}{4\pi\epsilon_{0}\hbar c} \frac{v}{c} \right)^{2}\! \left(\frac{\hbar v}{\lambda_{B}\Omega}\right)^{2} \!\!\left[ \sum_{n=1,2}\!2\gamma(\omega\! - \!\epsilon_{(n+1)^{+}})\! + \!\sum_{n\geq 3}\!\gamma(\omega\! - \!\epsilon_{(n+1)^{+}}\! - \!\epsilon_{(n-1)^{+}}) \!\right]\!\!. \nonumber
\end{align}\end{widetext}
Here, we use Lorentzian $\gamma(x)=\pi^{-1}\Gamma/(x^{2}+\Gamma^{2})$ with a width specified by $\Gamma$ to model the broadening of Landau levels. The term $\delta\!g(\omega)$ describes the spectral density of the $(n\mp 1)^{-}\!\rightarrow\! {(n\pm 1)}^{+}$ transitions, which is a correction to the dominant contribution due to the $n^{-}\!\rightarrow\! n^{+}$ transitions given by the first term on the right hand side of Eq. (\ref{raman_mag_field}).

An example of the low-energy electronic contribution to the Raman spectrum in the neutral bilayer in strong magnetic field is shown with a solid line in Fig. 2(b). The dominant features are peaks due to the $n^{-}\!\rightarrow\! n^{+}$ transitions with the first being the $2^{-}\!\rightarrow\! 2^{+}$ transition. Note that within the LL indexing scheme applied here, indices $0$ and $1$ are only used to denote one valley-degenerate level each (no $\alpha$ index is needed). Lifting the valley degeneracy by introducing charge asymmetry between layers will not allow any additional $n^{-}\!\rightarrow\! n^{+}$ transition because valley-split levels for $n=0,1$ belong to different valleys and excitation between them would require a huge momentum transfer. The quantum efficiency of a single $n^{-}\!\rightarrow\! n^{+}$ peak in Fig. 2(b) is approximately 
\begin{equation}
I_{n^{-}\rightarrow n^{+}}\approx \left( \frac{v^{2}}{c^{2}}\frac{e^{2}/\lambda_{B}}{\epsilon_{0}\pi\Omega} \right)^{2} =\frac{v^{4}e^{5}B}{\pi^{2}c^{4}\epsilon_{0}^{2}\hbar\Omega^{2}}
\end{equation}
per incoming photon, which at the field $B\sim 10$T gives $I_{n^{-}\rightarrow n^{+}}\sim 10^{-12}$ for $\Omega\sim 1$eV photons, comparable to similar transitions in monolayer graphene.\cite{kashuba_prb_2009}

\begin{figure}[btp]
\centering
\includegraphics[width=.98\columnwidth]{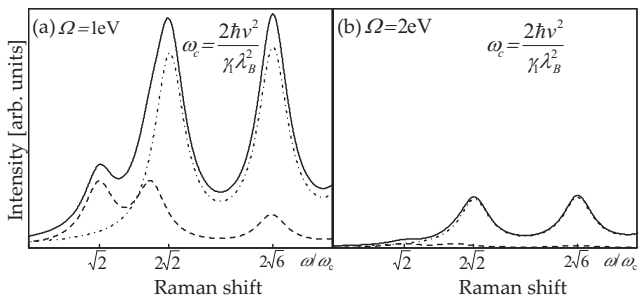}
\caption{Comparison of electronic contributions to the Raman spectra in neutral bilayer graphene for two different energies of incoming photons: (a) $\Omega=1$eV, and (b) $\Omega=2$eV. For each case, total spectral density $g(\omega)$ and contributions due to the $n^{-}\!\rightarrow\! n^{+}$ and $(n\pm 1)^{-}\!\rightarrow\! (n\mp 1)^{+}$ modes are shown in the solid, dot-dashed, and dashed line, respectively. Intensity scale is the same on (a) and (b); values of the parameters used: $v=10^{6}$m/s, $\gamma_{1}=0.4$eV, $B=10$T, and $\Gamma=0.012$eV. }
\label{f3}
\end{figure}

A weaker feature in Fig. 2(b) is the first and the only visible $(n\mp 1)^{-}\!\rightarrow\! {(n\pm 1)}^{+}$ peak due to both $2^{-}\rightarrow 0$ and $0\rightarrow 2^{+}$ transitions, positioned to the left of the $2^{-}\!\rightarrow\! 2^{+}$ peak. The quantum efficiencies of the $(n\pm 1)^{-}\!\rightarrow\! (n\mp 1)^{+}$ transitions are smaller by the factor $\left(\frac{\gamma_{1}}{\Omega}\right)^{2}$ in comparison to the $n^{-}\!\rightarrow\! n^{+}$ transitions. This is different from the monolayer graphene case, where the corresponding ratio between quantum efficiencies of $(n\pm 1)^{-}\!\rightarrow\! (n\mp 1)^{+}$ and $n^{-}\!\rightarrow\! n^{+}$ transitions is $\left(\frac{\omega}{\Omega}\right)^{2}$, much smaller than for the bilayer. The term $\delta\!g(\omega)$ can be further emphasized by changing the energy of incoming photons $\Omega$. Shown in Figs. 3(a) and 3(b), is a comparison of the total spectral density $g(\omega)$ and contributions due to each mode separately, for two different energies of incoming photons, $\Omega=2$eV and $\Omega=1$eV. The intensity scale is the same on both figures and in each case, the total spectral density $g(\omega)$, the contributions due to the $n^{-}\!\rightarrow\! n^{+}$ and $(n\pm 1)^{-}\!\rightarrow\! (n\mp 1)^{+}$ modes are shown in the solid, dot-dashed and dashed line, respectively. The dominant contribution, resulting from the Raman scattering accompanied by the $n^{-}\!\rightarrow\! n^{+}$ electronic transitions, is proportional to the inverse square of the incoming photon energy $\Omega$. Therefore, two peaks drawn with dot-dashed lines are roughly four times smaller on the right figure. The spectral density of the $(n\pm 1)^{-}\!\rightarrow\! (n\mp 1)^{+}$ transitions is smaller by a further factor $\left(\frac{\gamma_{1}}{\Omega}\right)^{2}$ in comparison to the $n^{-}\!\rightarrow\! n^{+}$ transitions. Hence, this contribution, shown with dashed lines, is close to zero on the right figure, while on the left, the first of the two smaller peaks corresponding to symmetric transitions  $2^{-}\rightarrow 0$, $0\rightarrow 2^{+}$ and $3^{-}\rightarrow 1$, $1\rightarrow 3^{+}$ is still visible in the total spectral density. Because of the contrasting polarization factors in Eq. (\ref{raman_mag_field}), contributions of different modes, $n^{-}\!\rightarrow\! n^{+}$ or $(n\pm 1)^{-}\!\rightarrow\! (n\mp 1)^{+}$, to the total spectral density could be separated using polarizers. If the polarizers were set as to collect only photons with circular polarization identical to that of the incoming photons, then the $n^{-}\!\rightarrow\! n^{+}$ contribution would be measured. However, if only the photons with polarization opposite to the polarization of the incoming beam were detected, the $(n\mp 1)^{-}\!\rightarrow\! {(n\pm 1)}^{+}$ contribution would be determined.

Increasing the filling factor leads first to the $2^{-}\rightarrow 0$ and $3^{-}\rightarrow 1$ transitions being blocked when LLs with $n=0$ and $n=1$ are completely filled. Therefore, the height of the two corresponding $(n\pm 1)^{-}\!\rightarrow\! (n\mp 1)^{+}$ peaks is halved (transitions $0\rightarrow 2^{+}$ and $1\rightarrow 3^{+}$ are still allowed). Next to disappear are the first $n^{-}\!\rightarrow\! n^{+}$ peak, that is $2^{-}\!\rightarrow\! 2^{+}$, and the remains of the first $(n\pm 1)^{-}\!\rightarrow\! (n\mp 1)^{+}$ peak, (due to the $0\rightarrow 2^{+}$ transition) because of the filled LL $2^{+}$. Complete filling of each following Landau level results in the disappearance of the next $n^{-}\!\rightarrow\! n^{+}$ and $(n\pm 1)^{-}\!\rightarrow\! (n\mp 1)^{+}$ peaks.

\section{summary}

We presented a theory of inelastic scattering of photons in bilayer graphene accompanied by the excitation of electron-hole pairs . Similar to monolayer graphene, the dominant scattering processes lead to the crossed polarisation of in/out photons. Also, the selection rules in the presence of a magnetic field are found to be the same, with the $n^{-}\!\rightarrow\! n^{+}$ mode being the strongest. We estimate the intensity of one of the $n^{-}\!\rightarrow\! n^{+}$ scattering processes to be $I_{n^{-}\rightarrow n^{+}}\sim 10^{-12}$ for $\Omega\sim 1$eV photons in magnetic field $B\sim 10$T. The most recent theoretical prediction for monolayer graphene of the intensity of the phonon-induced $G$ peak,\cite{basko_njp_2009} a well known Raman feature in carbon materials,\cite{ferrari_ssc_2007} estimates $I_{G}\sim 10^{-11}$. This result is only one order of magnitude greater than the intensity of a single $n^{-}\!\rightarrow\! n^{+}$ peak. Therefore, spectral features of inter-Landau-level transitions in bilayer graphene predicted in this paper may be observable experimentally.

The electron Raman scattering in an external magnetic field would complement infrared spectroscopy as it can give information about electronic excitations between different pairs of Landau levels. The purpose of this paper was to identify and describe the dominant inter-Landau-level modes. However, additional corrections e.g., to the Raman-active magneto-exciton energies, due to the many-body effects neglected in the above considerations will be present in the spectra. Many-body corrections were, for example, observed in the infrared spectroscopy experiment performed in external magnetic fields. \cite{henriksen_prl_2008} Electronic Raman measurements could provide a test ground for some of the theoretical models proposed to account for these many-body effects.

\begin{acknowledgments}

This project has been funded by the EPSRC grants EP/G041954 and Science \& Innovation Award EP/G035954.

\end{acknowledgments}

\appendix*

\section{The contact interaction}

The contact interaction scattering amplitude $\delta\!\mathcal{R}$, illustrated using Feynman diagram in Fig. 1(d), results from the second term in the interaction Hamiltonian (\ref{int_Hamiltonian}) and corresponds to the one-step Raman processes. It is characterised by operators $\partial^{2}{\hat{H}}/\partial p_{i}\partial p_{j}$ and hence, the bilayer graphene Hamiltonian in (\ref{4b_Hamiltonian}) does not allow any contact interaction processes, as it includes only terms linear in the electron momentum $\mathbf{p}$. The only contribution to the contact interaction comes from the addition $\delta\!{\hat{H}}$, which contains terms quadratic in $\mathbf{p}$. However, such contribution involves prefactor $\frac{v^{2}}{\gamma_{0}\Omega}$ and therefore leads only to small corrections in the intensity of Raman scattering of photons with energy less than the band-width of graphene, $\sim 6\gamma_{0}$. In fact, the contribution to the scattering amplitude due to the contact interaction obtained within the four-band model is
\begin{align}
& \delta\!\mathcal{R} = \frac{e^{2}\hbar^{2}v^{2}}{6\epsilon_{0}\Omega\gamma_{0}} \bm{\mathcal{L}}\!\cdot\!\mathbf{d};\;\;\bm{\mathcal{L}}=\left(\mathcal{L}_{x},\mathcal{L}_{y}\right); \\ & \mathcal{L}_{x} = \left( \begin{array}{cc}
-\frac{v_{3}}{v}\sigma_{y} & \sigma_{y} \\
\sigma_{y} & 0
\end{array} \right);\;
\mathcal{L}_{y} = \left( \begin{array}{cc}
-\frac{v_{3}}{v}\sigma_{x} & -\sigma_{x} \\
-\sigma_{x} & 0
\end{array} \right) \nonumber.
\end{align}
However, as we are interested only in the low-energy physics (small Raman shifts), only the top left block of the above matrix is relevant. This block contains an additional prefactor $\frac{v_{3}}{v}$. We see that the contact interaction can be neglected in comparison to the leading terms in the effective scattering amplitude $\mathcal{R}_{\textrm{eff}}$, as $\frac{v_{3}/v}{6\gamma_{0}}\ll\frac{\gamma_{1}}{\Omega^{2}}<\frac{1}{\Omega}$.



\begin{thebibliography}{99}

\bibitem{mccann_prl_2006} E. McCann and V.I. Fal'ko, Phys. Rev. Lett. {\bf 96}, 086805 (2006).

\bibitem{geim_novoselov_review_2007} A.K. Geim and K.S. Novoselov, Nature Materials {\bf 6}, 183 (2007).

\bibitem{mccann_prb_2006} E. McCann, Phys. Rev. B {\bf 74}, 161403(R) (2006).

\bibitem{oostinga_nature_materials_2007} J.B. Oostinga, H.B. Heersche, X. Liu, A.F. Morpurgo, and L.M.K. Vandersypen, Nat. Mat. {\bf 7}, 151 (2008).

\bibitem{ohta_science_2006} T. Ohta, A. Bostwick, T. Seyller, K. Horn, and E. Rotenberg, Science {\bf 313}, 951 (2006).

\bibitem{mucha-kruczynski_prb_2008} M. Mucha-Kruczy\'{n}ski, O. Tsyplyatyev, A. Grishin, E. McCann, V.I. Fal'ko, A. Bostwick, and E. Rotenberg, Phys. Rev. B {\bf 77}, 195403 (2008).

\bibitem{nair_science_2008} R.R. Nair, P. Blake, A.N. Grigorenko, K.S. Novoselov, T.J. Booth, T. Stauber, N.M.R. Peres, and A.K. Geim, Science {\bf 320}, 1308 (2008).

\bibitem{gaskell_apl_2009} P.E. Gaskell, H.S. Skulason, C. Rodenchuk, and T. Szkopek, Appl. Phys. Lett. {\bf 94}, 143101 (2009).

\bibitem{zhang_prb_2008} L.M. Zhang, Z.Q. Li, D.N. Basov, M.M. Fogler, Z. Hao, and M.C. Martin, Phys. Rev. B {\bf 78}, 235408 (2008).

\bibitem{kuzmenko_prb_2009a} A.B. Kuzmenko, E. van Heumen, D. van der Marel, P. Lerch, P. Blake, K.S. Novoselov, and A.K. Geim, Phys. Rev. B {\bf 79}, 115441 (2009).

\bibitem{kuzmenko_prb_2009b} A.B. Kuzmenko, I. Crassee, D. van der Marel, P. Blake, and K.S. Novoselov, Phys. Rev B {\bf 80}, 165406 (2009).

\bibitem{li_prl_2009} Z.Q. Li, E.A. Henriksen, Z. Jiang, Z. Hao, M.C. Martin, P. Kim, H.L. Stormer, and D.N. Basov, Phys. Rev. Lett. {\bf 102}, 037403 (2009).

\bibitem{mak_prl_2009} K.F. Mak, C.H. Lui, J. Shan, and T.F. Heinz, Phys. Rev. Lett. {\bf 102}, 256405 (2009).

\bibitem{kuzmenko_prl_2009} A.B. Kuzmenko, L. Benfatto, E. Cappelluti, I. Crassee, D. van der Marel, P. Blake, K.S. Novoselov, and A.K. Geim, Phys. Rev. Lett. {\bf 103}, 116804 (2009).

\bibitem{ferrari_prl_2006} A.C. Ferrari, J.C. Meyer, V. Scardaci, C. Casiraghi, M. Lazzeri, F. Mauri, S. Piscanec, D. Jiang, K.S. Novoselov, S. Roth, and A.K. Geim, Phys. Rev. Lett. 97, 187401 (2006).

\bibitem{gupta_nano_letters_2006} A. Gupta, G. Chen, P. Joshi, S. Tadigadapa, and P.C. Eklund, Nano Lett., {\bf 6}, 2667 (2006).

\bibitem{graf_nano_letters_2007} D. Graf, F. Molitor, K. Ensslin, C. Stampfer, A. Jungen, C. Hierold, and L. Wirtz, Nano Lett. {\bf 7}, 238 (2007).

\bibitem{malard_prb_2007} L.M. Malard, J. Nilsson, D.C. Elias, J.C. Brant, F. Plentz, E.S. Alves, A.H. Castro Neto, and M.A. Pimenta, Phys. Rev. B {\bf 76}, 201401(R) (2007).

\bibitem{yan_prl_2008} J. Yan, E.A. Henriksen, P. Kim, and A. Pinczuk, Phys. Rev. Lett. {\bf 101}, 136804 (2008).

\bibitem{malard_prl_2008} L.M. Malard, D.C. Elias, E.S. Alves, and M.A. Pimenta, Phys. Rev. Lett. {\bf 101}, 257401 (2008).

\bibitem{das_prb_2009} A. Das, B. Chakraborty, S. Piscanec, S. Pisana, A.K. Sood, and A.C. Ferrari, Phys. Rev. B {\bf 79}, 155417 (2009).

\bibitem{ni_prb_2009} Z. Ni, L. Liu, Y. Wang, Z. Zheng, L.-J. Li, T. Yu, and Z. Shen, Phys. Rev. B {\bf 80}, 125404 (2009).

\bibitem{mafra_prb_2009} D.L. Mafra, L.M. Malard, S.K. Doorn, H. Htoon, J. Nilsson, A.H. Castro Neto, and M.A. Pimenta, Phys. Rev. B {\bf 80}, 241414(R) (2009).

\bibitem{abstreiter_springer_1984} G. Abstreiter, M. Cardona, and A. Pinczuk in {\it Light Scattering in Solids IV}, edited by M. Cardona and G. Guntherodt (Springer, Berlin, 1984).

\bibitem{klein_springer_1983} M.V. Klein in {\it Light Scattering in Solids I}, edited by M. Cardona (Springer, Berlin, 1983).

\bibitem{wolff_prl_1966} P.A. Wolff, Phys. Rev. Lett. {\bf 16}, 225 (1966).

\bibitem{slusher_prl_1967} R.E. Slusher, C.K.N. Patel, and P.A. Fleury, Phys. Rev. Lett. {\bf 18}, 77 (1967).

\bibitem{patel_prl_1968} C.K.N. Patel and R.E. Slusher, Phys. Rev. Lett. {\bf 21}, 1563 (1968).

\bibitem{kashuba_prb_2009} O. Kashuba and V.I. Fal'ko, Phys. Rev. B {\bf 80}, 241404(R) (2009).

\bibitem{abergel_prb_2007} D.S.L. Abergel and V.I. Fal'ko, Phys. Rev. B {\bf 75}, 155430 (2007).

\bibitem{henriksen_prl_2008} E.A. Henriksen, Z. Jiang, L.-C. Tung, M.E. Schwartz, M. Takita, Y.-J. Wang, P. Kim, and H.L. Stormer, Phys. Rev. Lett. {\bf 100}, 087403 (2008).

\bibitem{dresselhaus_adv_physics_1981} M.S. Dresselhaus and G. Dresselhaus, Adv. Phys. {\bf 30}, 139 (1981).

\bibitem{footnote_v3} In the integration over the electronic momentum $\mathbf{p}$ we neglected the trigonal warping of the electronic dispersion caused by $v_{3}$. This effect is important only for very low energies.

\bibitem{abergel_epjst_2007} D.S.L. Abergel, E. McCann, and V.I. Fal'ko, Eur. Phys. J. Spec. Top. {\bf 148}, 105 (2007).

\bibitem{mucha-kruczynski_jpcm_2009} M. Mucha-Kruczy\'{n}ski, D.S.L. Abergel, E. McCann, and V.I. Fal'ko, J. Phys.: Condens. Matter {\bf 21}, 344206 (2009).

\bibitem{basko_njp_2009} D.M. Basko, New J. Phys. {\bf 11}, 095011 (2009).

\bibitem{ferrari_ssc_2007} A.C. Ferrari, Solid State Commun. {\bf 143}, 47 (2007).

\end{thebibliography}
\end{document}